# High Frequency Microwave Emission of a Tri-layer Magnetic Tunnel Junction in the Absence of External Bias-Magnetic Field

R.N.S. Rajapakse[†], Z. M. Zeng[‡], H.W. Jiang[†*]

[†]*Department of Physics and Astronomy, University of California Los Angeles, USA.*

[‡]*Key Laboratory of Multifunctional Nanomaterials and Smart Systems, Suzhou Institute of Nano-tech and Nano-bionics, Chinese Academy of Sciences, Ruoshui Road 398, Suzhou 215123, P. R. China*

*Corresponding author - H.W. Jiang (jiangh@physics.ucla.edu)*

## Abstract

We perform an experimental study of DC current induced microwave emission in a magnetic tunnel junction (MTJ) consisting of three active magnetic layers. For this tri-layer structure, in addition to a conventional bilayer orthogonal MTJ containing a perpendicular free layer and an in-plane fixed layer, a second perpendicular layer has been introduced. We found that the microwave emission frequency induced by spin-transfer torque (STT) can reach as high as 6 GHz in the absence of any applied magnetic field. Moreover, microwave emission is observed for both current polarizations where a redshift is seen with increase in magnitude of current. We discuss spin-dynamics of the observed bi-directional high-frequency emission and the physical origin of the red-shift. The distinct microwave emission properties exhibited in this tri-layer MTJ structure could potentially be useful for future applications in nanoscale spintronics devices such as microwave communication and neuromorphic computing.

Injection of spin polarized current into a nano sized magnet can excite steady state magnetization precessions via STT[1,2] and with tunneling magnetoresistance (TMR) this can be captured as an electrical signal measurable at the electrodes. Spin transfer torque nano oscillators (STNO)[3–6] utilizing these phenomena have drawn great attention as future nanoscale microwave devices due to its scalability, high frequency tunability, broad working temperature and high compatibility with the existing semiconductor manufacturing processes[7–10]. These promising devices have developed tremendously over the past years in many aspects for example, enhancing output power[11–14] and improving linewidth[15–18]. One of the major attempts has been to eliminate the need of an external bias field[19–23] but still achieving reasonable output microwave frequency in the interest of practical applications. Recently, STNOs have drawn even more attention as building blocks for neural networks in computational tasks for artificial intelligence[24–27]. These emerging new avenues for STNOs in timely research interests have further proved their promising role in future technology.

Commonly, STNOs are formed with a pinned polarization layer and a free precessing layer where the STT induced emission can be obtained only for one current direction[5]. We report STT induced microwave emission in an MTJ consisting of three active magnetic layers where an additional perpendicular layer is introduced as the third on the top of the otherwise conventional stack. We found the emission frequency can reach as high as 6 GHz in the absence of any external field due to the enhanced interfacial perpendicular magnetic anisotropy[28,29] (IPMA) in the middle layer owing to the additional CoFeB/MgO interface introduced along with the top layer. Another unique feature of this STNO is STT microwave emission can be realized for both current polarizations due to dual polarizer action of top and bottom layers. Our experiments in magnetic fields have revealed the spin dynamics of the oscillator as out-of-plane (OOP) modes induced in the middle layer. These new functionalities of the tri-layer STNO could potentially electrify array of applications such as microwave communication, neuromorphic computing, artificial intelligence *etc*.

# Results

## Device Characterization

The structure of the MTJ presented in this paper is substrate/buffer-layer/PtMn(20)/Co$_{70}$Fe$_{30}$(2.3)/Ru(0.85)/Co$_{60}$Fe$_{20}$B$_{20}$(2.7)/MgO(0.8)/Co$_{20}$Fe$_{60}$B$_{20}$(1.4)/MgO(0.85)/Co$_{20}$Fe$_{60}$B$_{20}$(1.5)/capping-layer where thicknesses within the parenthesis are in nanometers. As shown in Figure 1a, this can be understood as three principle magnetic layers referred to as top, middle and bottom where the bottom layer is the top section of a synthetic antiferromagnetic layer[30] usually referred as "reference layer". The Fe-rich CoFeB in the middle and top layers are used to create stronger perpendicular magnetic anisotropy (PMA) in the CoFeB/MgO interface[29]. The two tunnel junctions corresponding to each MgO barrier form two resistors that are connected in series as shown on the right-hand-side of figure 1a therefore, the total resistance of the MTJ is the summation of these individual resistance values. In this study, positive current is defined to be electrons flowing from bottom to the top layer. The convention of the coordinate system for optional external magnetic field is as shown in figure 1a. Perpendicular magnetoresistance ($MR_Z$) loop obtained at 10 µA is demonstrated in figure 1b where insets illustrate individual magnetization configurations at different points. Two piecewise linear segments are present in between which two hysteresis transitions occur. Low TMR variation seen in both linear segments and transitions suggests tri-layer magnetizations maintain close mutual angles throughout the $MR_Z$ loop including the major reorientation (switching).

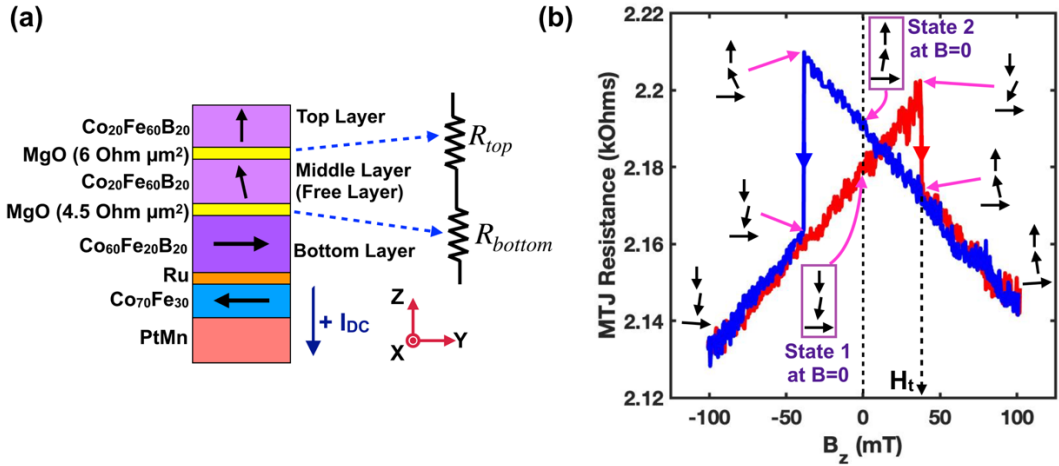

Figure 1. MTJ structure and characterization. (a) Schematic diagram of the MTJ at zero bias field. Left represents the key layers of the design while the right represents the corresponding electrical circuit model. The device consists of three layers referred to as top, middle and bottom. Top and middle layers possess perpendicular easy axis while bottom possesses in-plane easy axis. Convention of the coordinate system and the positive current are also shown. (b) Magnetoresistance curve for a fixed bias current (10 µA) for perpendicular magnetic field. Insets illustrate magnetization orientations of the three layers explaining linear segments and hysteresis transitions. Red and blue curves correspond to field sweeping up and down respectively. As shown, zero-field configuration can be initialized in to either state 1 or 2 with applying a negative or positive large field. The definition of right-hand transition field, $H_t$ is shown.

Top and middle layer thicknesses are well below the threshold required to induce a perpendicular easy axis via IPMA[29] while the bottom layer is a well-established in-plane reference layer[30–32]. Within the studied field range (150 mT for $B_Z$), the bottom is expected to be affected minimally by the external field. Top layer is considered to be normal to the film plane while the middle is expected to have canted out from the vertical axis due to the dipole coupling from the bottom. In a conventional bi-layer MTJ, such a canting angle is about[33] 10º however, that here is expected to be much less due to high IPMA field from dual CoFeB/MgO interfaces and the vertical dipole coupling from the top layer. (This canting angle is exaggerated in figures for clarity). As shown in figure 1b, MTJ's zero field configuration can be initialized in two different states namely, state-1 and state-2. Here, state-1 (state-2) is the configuration with both top and middle magnetizations prepared to point downwards (upwards). State-1 (state-2) can be realized by applying a large negative (positive) $B_Z$ prior to the experiment. Observed linear segments result from minor variations of middle layer canting angle while hysteresis transitions are due to switching of top and middle layers. The field at which the right hand side transition occurs is defined as $H_t$ (Figure 1b) for later reference. $MR_X$ and $MR_Y$ loops have also been obtained. (see supporting information).

**Microwave Emission Measurements**

Microwave emission in the absence of any external field is recorded for both current directions with simultaneous resistance measurements. Figure 2a and b represent the recorded emission, calculated power delivered to a matched load and simultaneous device resistance respectively. Figure. 2c and d present individual spectra for select positive and negative current values respectively where the curves are vertically shifted by 40 nWGHz$^{-1}$ for clarity. Emission spectrum resembles an inverted parabolic shape slightly shifted towards the positive current direction. Emission builds up for both ±$I_{DC}$ with a redshift up to a certain point beyond which highly incoherent oscillations start to occur as indicated by the sudden broadening in the emission peak. Within the region of coherent oscillations, emission power has increased beyond 90 nW for both current directions which is about two orders of magnitude greater than what thermally induced emission can offer[34] for MgO based MTJs, suggesting the occurrence of STT induced oscillations for both current directions. Furthermore, within this coherent emission region for both $\pm I_{DC}$, device resistance has dropped by more than 15% of its zero-field value (Figure 2b) revealing a considerable tilting of the precessing magnetization away from its equilibrium axis. This fact further validates the prediction that emission (for both $\pm I_{DC}$) is STT induced rather than thermal induced where such a drop in resistance cannot be justified.

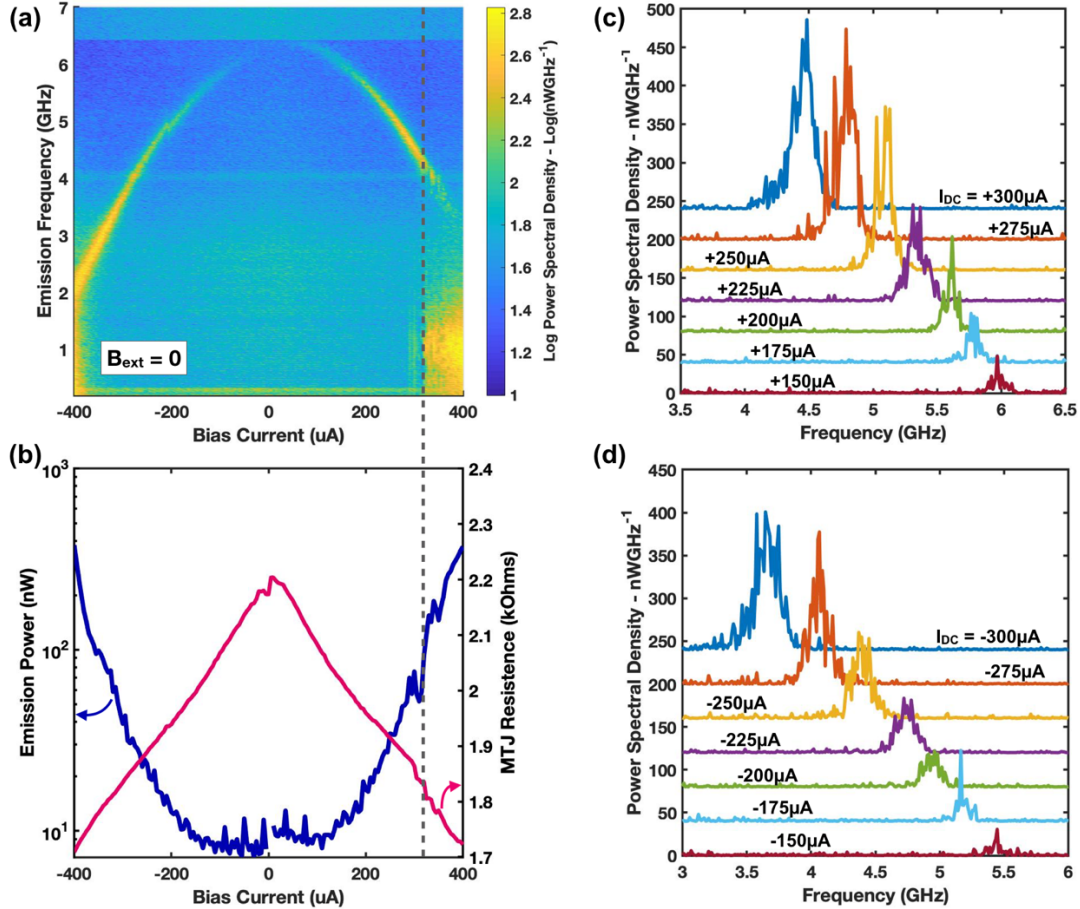

Figure 2. Microwave emission at zero bias-field. (a) Microwave emission spectrum for both current directions obtained at zero external magnetic field. Staring from about 6 GHz a redshift is observed for both current polarities where beyond +320 µA and –400 µA emission has become highly incoherent. (b) Integrated power delivered to a matched load (blue and left axis) and simultaneous resistance measured during experiment (pink and right axis). Dash-line is as a guide to identify simultaneous footprints at the onset of incoherent emission on the positive current direction. As seen resistance has changed by more than 15% within the coherent oscillation region for both current polarities. (c) and (d) Emission spectra for select positive and negative bias current values respectively where the curves are vertically offset by 40 nWGHz$^{-1}$ for clarity.

For a perpendicular precessing layer, oscillating frequency is directly proportional to the local effective field inside[5]. Having the two sides of the emission parabola smoothly connected across zero current suggests it is the same layer (either top or middle) producing emission for both $\pm I_{DC}$. To further investigate into the oscillating layer and its modes, device's emission response to an external field has been studied. An important clarification to be made here is, the purpose of this external field is to observe its effect on the precessional cone and emission frequency which is otherwise capable of producing emission at zero field. An interesting distinction in the response is observed between state-1 and state-2 initializations. Results recorded at a fixed bias of +150 µA are presented in figure 3a and b for state-1 and state-2 initializations respectively.

Upon state-1 initialization, as $B_Z$ is swept in positive direction, a linear redshift is observed up to a critical field where a discontinuous transition occurs to a higher frequency which then followed by a continuous linear blueshift. In contrast, for state-2 initialization only the blueshift is present from the very beginning. Apart from this, the power is observed to increase along the redshift but to decrease along the blueshift. It turns out that both of these observations can be explained with OOP oscillations induced in a perpendicular free layer of which the precession cone evolves as illustrated in the insets of figures 3a and 3b. Here the axis of the cone has been taken to be vertical for simplicity of illustration. Given both top and middle layers are perpendicular, the question is which out of those two is in fact producing the observed microwave output. The possibility of top layer producing the observed emission is eliminated due to lack of an appropriate neighboring reference layer. In contrast, for the middle layer, in-plane bottom layer is present which can successfully capture the dynamical TMR during OOP precessions in the former.

Considering OOP precessional modes in the middle layer, observations in figure 3 can be explained as follows. After state-1 initialization, with increasing upward field, downward oriented cone expands up to a point after which it abruptly reorients to be upwards. Prior to this reorientation, magnitude of effective field inside the middle layer is $\mu_o(H_k - M_S)m_z - B_{ext}$ where $H_k$, $M_S$, $m_z$ and $B_{ext}$ are IPMA field, saturation magnetization, vertical fractional magnetization and external field respectively. Following the reorientation, that is modified to $\mu_o(H_k - M_S)m_z + B_{ext}$. This discontinuity of the magnitude of the internal field through the transition can then explain the observed frequency jump. Furthermore, with the increase (decrease) of cone angle, vertical component of the effective field within the layer decreases (increases) which explains the redshift and blueshift within each branch. The emission power is directly proportional to the maximum dynamical resistance of the device during one precession[5] which indeed increases with increasing cone angle hence justifying the observed increase (decrease) in emission power along the redshift (blueshift) as the cone continues to enlarge (suppress) under the field. Following the reorientation, now being aligned with external field, cone will suddenly be quenched explaining the sudden drop in power as seen in figure 3a.

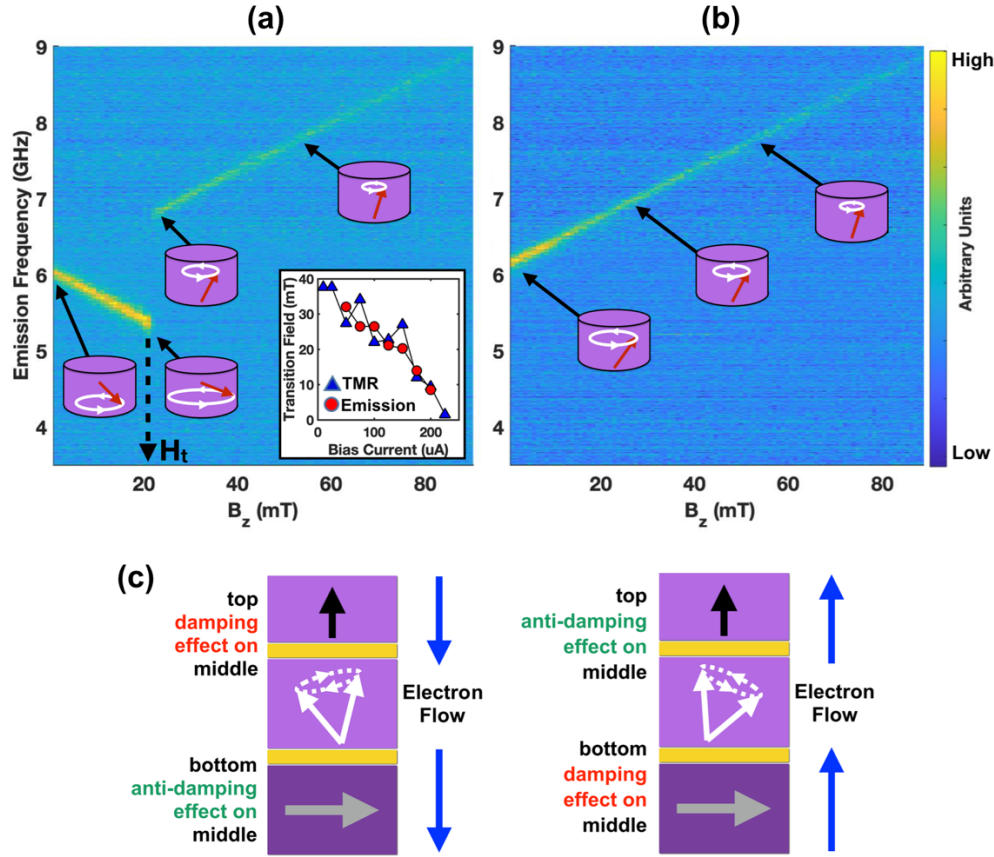

Figure 3. Oscillation modes in the middle layer. (a) Manipulation of the emission with perpendicular external field at a fixed bias current of +150 µA upon state-1 initialization. Definition of the transition field $H_t$ is shown. Insets of the precessing middle layer illustrate the evolution of the cone angle. In state-1, middle layer starts off to precess with a downward cone and the cone angle is enhanced as $+B_Z$ is increased along with a redshift. At the transition field ($H_t$) cone undergoes an abrupt reorientation to be upward causing a frequency jump in the spectrum followed by a blueshift. The inset graph shows the correlation between the transition fields at different bias currents derived from this emission cone flipping method and magnetoresistance measurements as defined in figure 1b. (b) Manipulation of emission upon state-2 initialization (bias current, +150 uA). In state-2, middle layer starts off to precess with an upward cone and the cone angle will be suppressed by $+B_Z$ with a blueshift. (c) Action of the top and bottom layers on the middle for each current direction. When electrons flow from top to the bottom layer (i.e. $-I_{DC}$), top layer produces a damping effect on the middle while the bottom layer produces an anti-damping effect on it. For the other current polarity (i.e. $+I_{DC}$) these roles played by top and bottom layers are switched.

For state-2 initialization, in contrast, oscillation cone is upwards to start with resulting in a continuous blueshift along with a decrease in power as the cone continues to suppress monotonically. The field at which the frequency discontinuity occurs for state-1 initialized case is defined as $H_t$ as shown in figure 3a. Occurrence of this discontinuity (in state-1) for different $I_{DC}$ have also been recorded (see supporting information). Furthermore, $MR_Z$ loops for different $I_{DC}$ are obtained (see supporting information). For both cases, corresponding $H_t$ values are derived as defined in figure 1b and figure 3a. It turns out that $H_t$ derived via both of these methods decrease with increase of $I_{DC}$ as, with increase of $I_{DC}$, STT rotates the magnetization further away from vertical axis hence facilitating the completion of the switching at a lower external field than otherwise. In fact it is found for $I_{DC}$ greater than 200 µA, only the continuous blueshift is present even upon state-1 initialization indicating the current itself has completed the switching without any aid of external field. Comparison of $H_t$ derived from these two independent methods demonstrates good correlation as seen in the inset of figure 3a, suggesting the specific event on figure 3a directly corresponds to the switching event on the $MR_Z$ loop further validating the OOP oscillation model.

## Discussion

One unique feature of the work presented here is the presence of STT induced emission for both current directions. In conventional bi-layer STNOs, for one current polarity,

STT acts as an anti-damping torque while for the other it acts as an additional damping torque[1,2], hence leaving room for persistent STT oscillations only for the former case given the threshold current magnitude is met. In our device, top and bottom layers are considered to have hybrid roles facilitating the observed bidirectional emission. Computational studies[35] on a bias-field free STNO with an in-plane polarizer and a perpendicular precessing layer have found, for parameters of a typical MTJ, bias-field free emission can be realized when the field-like torque (FLT) is present and its ratio with the STT attains a negative value. Given this scenario the perpendicular middle layer of the tri-layer device will experience a damping (anti-damping) STT by the bottom when electrons pass through the bottom (middle) layer to the middle (bottom) layer. In the meantime, the STT from the top layer will create a damping (anti-damping) effect on the middle when electrons pass through the top (middle) layer to the middle (top) layer. This is summarized in figure 3c. The strength of such damping and anti-damping torques do depend on the instantaneous magnetization direction and orientation of the cone itself which vary with the current direction. Therefore, with right quantitative conditions sustainable oscillations can be expected to occur owing to an overall net anti-damping effect for both current directions.

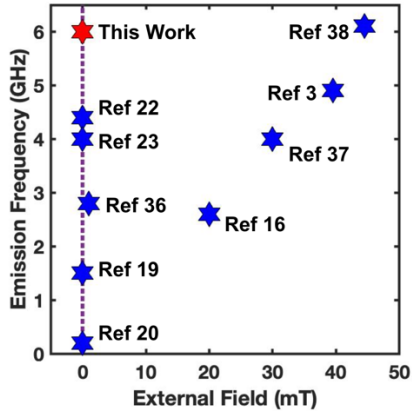

Figure 4. Comparison of work presented in this paper (red star) with existing work done at room temperature at zero bias field or below 50mT. The vertical line is a guide for the work done at zero field.

Another evidence for such dual polarizer action is the horizontal shift of emission parabola. This is a consequence of having different critical current values for each polarities which is justifiable with having different polarizers in action for each case. As seen in figure 2a, magnitude of the critical current for $+I_{DC}$ is greater than that for $-I_{DC}$ suggesting the polarizer in action for positive current is weaker in strength than the other. As illustrated in figure 3c, top (bottom) layer acts as the polarizer for $+I_{DC}$ ($-I_{DC}$). This is consistent with the above observation as, the bottom layer being a well-established polarizer, possesses higher polarization strength compared to the top layer.

Finally, Figure 4 puts our work in the context of some select achievements reported to date at room temperature and at zero bias field[19,20,22,23] or at a field less than 50 mT[3,16,36–38]. As it is seen the work reported here has the highest zero-field emission frequency to the best of our knowledge. All the results presented here are from a single device of 100 nm in diameter but qualitatively similar results have been observed in many other samples with different diameters and varying layer thicknesses. From a practical point of view the best interest of this work is the high frequency offered at zero-bias field. We attribute this to the dual MgO layers producing high IPMA field inside the middle layer which is estimated to be $\mu_o H_k \sim 1.6T$. Strength of IPMA from a CoFeB/MgO interface increases with decreasing CoFeB thickness however saturates below a certain thickness[29] which have been overcome in this study with the second such interface. Redshift observed for both $\pm I_{DC}$ results due to reduction in vertical component of the local effective field with the increase of precessing cone angle. Decrease of IPMA and $M_S$ due to Joule heating can be another cause to enhance the inherit redshift. Spin dynamics wise the other key feature observed is the presence of bi-directional STT induced emission. To accommodate this bi-directional emission in STT picture, dual roles played by top and bottom layers have been qualitatively discussed.

## Conclusion

In summary, we have demonstrated bi-directional STT induced microwave emission of a tri-layer MTJ that offers up to 6 GHz in the absence of any external field which is the highest achieved at zero-field to date to the best of our knowledge. Change in MTJ resistance and emission power have been considered to eliminate the possibility of thermally induced emission. We attribute this high frequency to the enhanced IPMA field owing to the dual CoFeB/MgO interfaces and bi-directional emission to the dual polarizer action of top and bottom layers. We hope this work will electrify the ongoing expedition of STNO based microwave communication and neuromorphic computing.

## Methods

In fabricating the devices, films were deposited using a Singulus TIMARIS PVD system and annealed at 300°C for 2.0 hours in a magnetic field of 1 T. Top and middle layers are deposited with varying thickness such that cross-sections forming perpendicular wedges allowing to select a device among variety of layer thickness combinations. The films were then patterned into nanopillars with different aspect ratios using electron-beam lithography and ion milling techniques. All results presented in this paper are obtained at room temperature and for a single circular device with diameter 100 nm. Electrical contacts with the device are made using a high bandwidth (40 GHz) microwave probe and emission is directly recorded at a 26 GHz spectrum analyzer through a bias-tee and an amplification of 28 dB. The complete experimental setup is present in the supporting information. Presented emission power (unless provided in arb. units) are corrected for this amplification and transmission inefficiency due to impedance mismatch between the device and the rest of the circuit.

## Acknowledgements

This work was supported by FAME, a STARnet center and the NSF under grant # DMR-1809155.

## Supporting Information

Experimental setup, All 3-axis magnetoresistance loops, $MR_Z$ loops at different $I_{DC}$, Evolution of state-1 initialized emission under $B_Z$ at different $I_{DC}$.

**Supporting Information**

**High Frequency Microwave Emission of a Tri-layer Magnetic Tunnel Junction in the Absence of External Bias-Field**

R.N.S Rajapakse[1], Z. Zeng[2], H.W. Jiang[1]

[1]*Department of Physics and Astronomy, University of California Los Angeles, USA.*

[2]*Key Laboratory of Nanodevices and Applications, Suzhou Institute of Nano-tech and Nano-bionics, Chinese Academy of Sciences, Ruoshui Road 398, Suzhou 215123, P. R. China*

*\*Corresponding author - H.W. Jiang ([jiangh@physics.ucla.edu](jiangh@physics.ucla.edu))*

## 1. Experimental setup

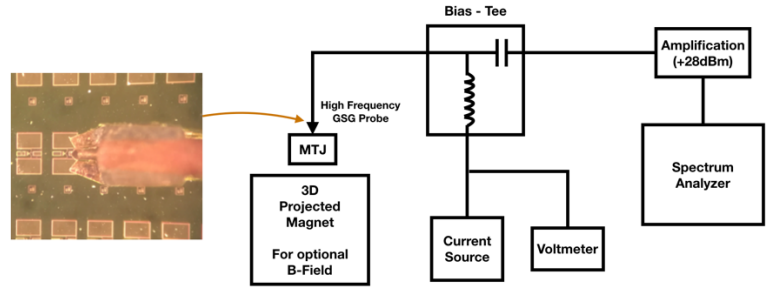

Figure S1 - Experimental setup. On the left is the view of the GSG probe making electrical contact with the device as observed through a microscope.

## 2. X,Y and Z magnetoresistance loops with comparison

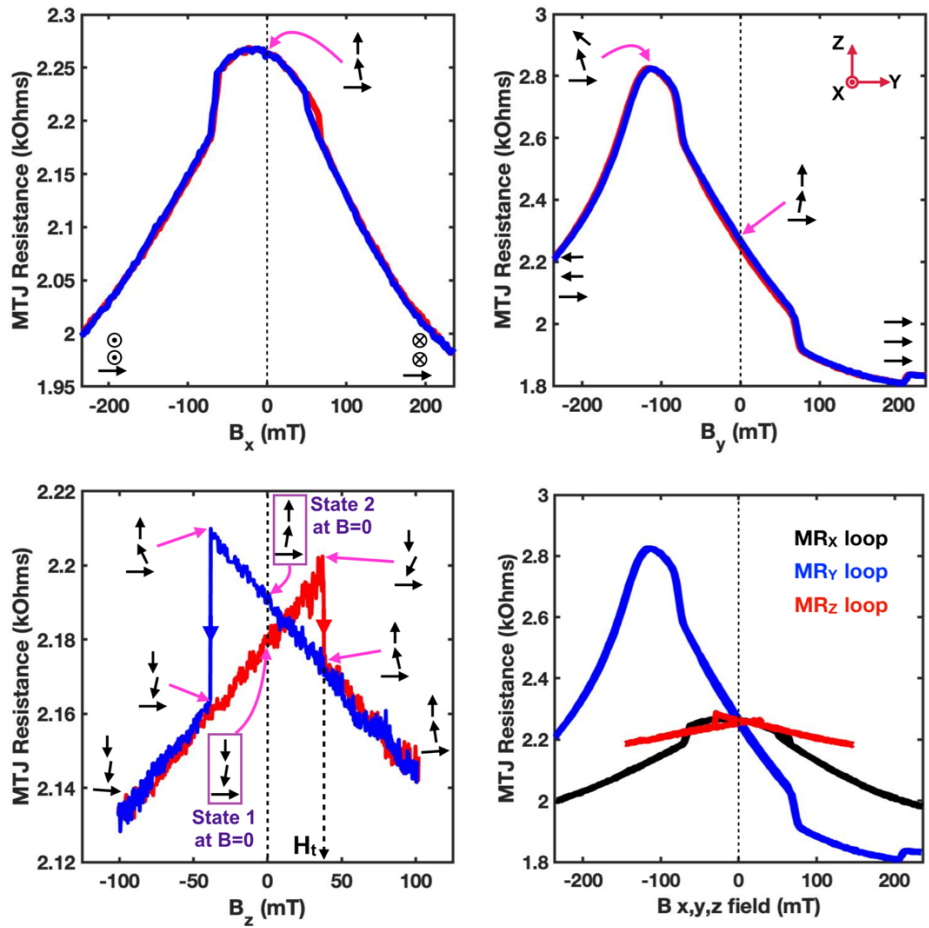

Figure S2 - $MR_X$, $MR_Y$, $MR_Z$ loops obtained at $I_{DC}= 10\mu A$.

All MR$_{X,Y,Z}$ loops have also been plot in one axis system for comparison. Both minimum and maximum resistance values are seen in MR$_Y$ loop, therefore bottom layer (or reference layer) is identified to be along +y direction.

### 3. MR$_Z$ loops at different bias current values

\

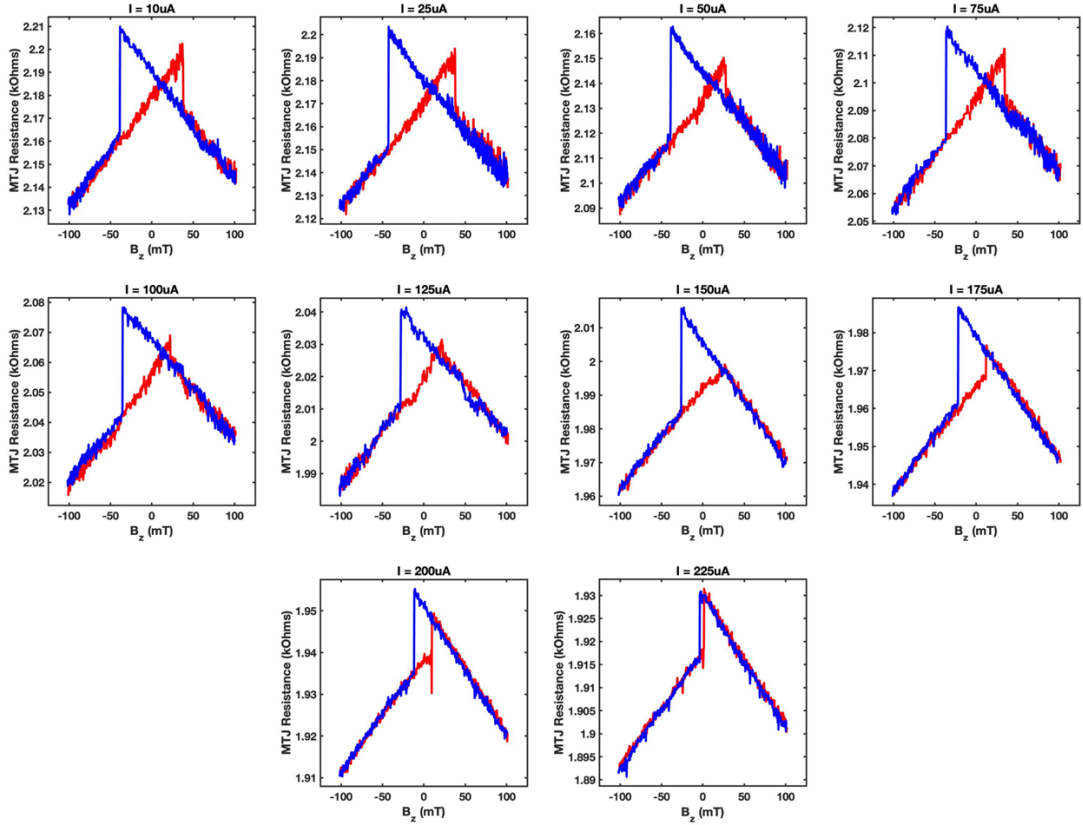

Figure S3 - MR$_Z$ loop at different I$_{DC}$.

MR$_Z$ loop for 10μA is presented in the main paper along with the definition of H$_t$, the field at which the right hand side hysteresis transition occurs. Above are MR$_Z$ loops obtained at different I$_{DC}$ up to 225μA. Qualitatively they all possess two linear segments in between which two hysteresis type transitions occur. Depending on the bias current value, the location (or the field of occurrence) of the transition changes mimicking triangles or trapezoids in the curve.

## 4. Emission under $B_Z$ at different $I_{DC}$ (for state-1 initialized)

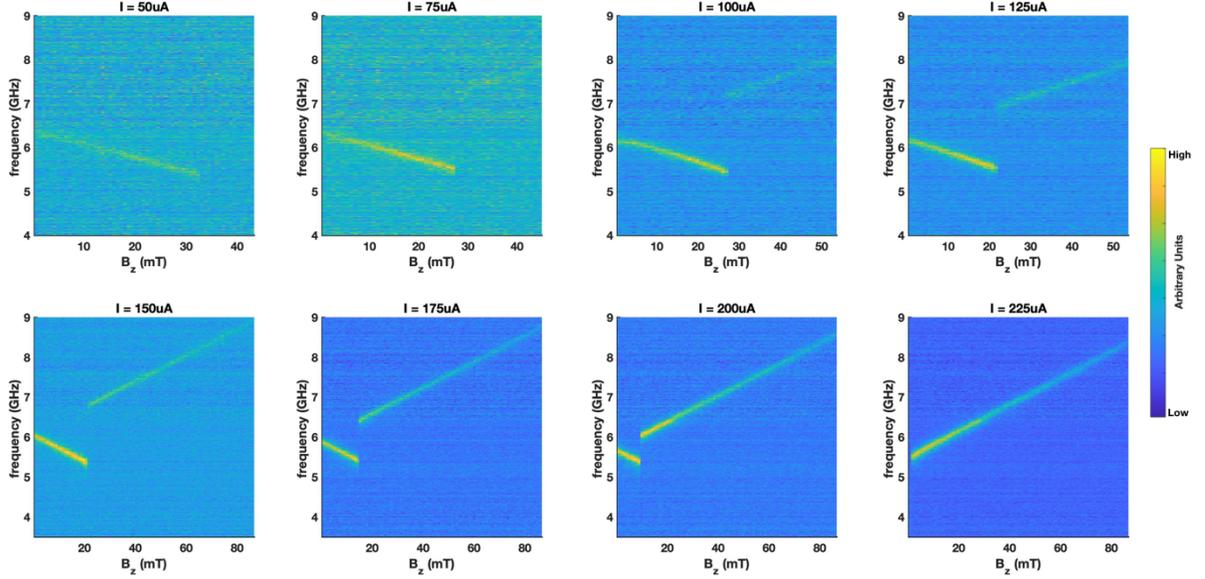

Figure S4 - Evolution of emission at different $I_{DC}$ upon state-1 initialization as $B_Z$ swept in the positive direction.

Evolution of the emission with sweeping $B_Z$ at $I_{DC}$ = 150μA (for both state-1 and state-2 initializations) are presented in the main paper along with the definition of $H_t$, the field at which the frequency discontinuity occurs for state-1 initialization. As It can be seen above, for state-1 initialization, when $I_{DC}$ is increased, $H_t$ decreases. At 225μA it is found that, even for state-1 initialization (both top and middle magnetizations prepared to point downwards), STT from the bias current itself has completed the switching (as indicative by presence of only the blueshift) without any aid of an external field. This agrees well with the observation that, as $I_{DC}$ approaches 225μA, width of the hysteresis of MR$_Z$ loop approaches zero.